\documentclass[epj]{svjour}
\usepackage{graphicx,bm,amsmath}
\newcommand{\up}{\uparrow}
\newcommand{\dn}{\downarrow}
\newcommand{\vect}{\bm}
\newcommand{\ep}{\epsilon}

\begin{document}
\title{Polaronic and dressed molecular states in orbital Feshbach resonances}

\author{
Junjun Xu\inst{1}\thanks{\email{jxu@ustb.edu.cn}}
\and
Ran Qi\inst{2}\thanks{\email{qiran@ruc.edu.cn}}}
\institute{
Department of Physics, University of Science and Technology Beijing, Beijing 100083, China
\and
Department of Physics, Renmin University of China, Beijing 100872, China}
\date{Received: date / Revised version: date}
%
\abstract{
We consider the impurity problem in an orbital Feshbach resonance (OFR), with a single excited clock state $|e\up\rangle$ atom immersed in a Fermi sea of electronic ground state $|g\dn\rangle$. We calculate the polaron effective mass and quasi-particle residue, as well as the polaron to molecule transition. By including one particle-hole excitation in the molecular state, we find significant correction to the transition point. This transition point moves toward the BCS side for increasing particle densities, which suggests that the corresponding many-body physics is similar to a narrow resonance.
}

\maketitle

\section{Introduction}
\label{intro}
The behaviours of a system with impurities are of particular interest in condensed matter physics. One of the famous examples is the Kondo effect, where the electric resistivity changes dramatically due to the magnetic impurities at low temperature \cite{Kondo}. Theoretically, we can consider an impurity model as minority spin-$\up$ particles immersed in a spin-$\dn$ Fermi sea. This system may be in a quasi-particle state, where each single impurity with the surrounding spin-$\dn$ particles forms a polaron; or it could be in a dressed molecular state, where the impurity could form cooper pairs with the spin-$\dn$ fermions.

Such a simplified model is especially realizable in cold atomic systems, due to the newly development of laser cooling and trapping\cite{review1,review2}. Furthermore, the realization of magnetic Feshbach resonances allows one to study the crossover from Bose-Einstein condensation (BEC) side to Bardeen-Cooper-Schrieffer (BCS) side continuously \cite{Feshbach1,Feshbach2}. On the deep BCS side, the ground state is the polaronic state, while the dressed molecule with pairing around Fermi energy is energetically not favarable. On the deep BEC side, the molecule is deeply bounded and becomes the ground state. This impurity related problems have been studied in various experimental and theoretical researches \cite{polaron1,polaron2,polaron3,polaron4,polaron5,1PH1,1PH2,1PH3,2D}.

Recent studies of many-body physics across narrow Feshbach rsonances have revealed rich new physics due to the strong energy dependence in scattering amplitude\cite{Ho,Hara}. For example, the polaron to molecule transition will move to the BCS side for sufficient narrow resonance width \cite{polaron2,Qi,Castin,Massignan}. However, most narrow resonances discovered in alkli atoms is difficult to control due to the smallness of $\Delta B$ (i.e. the resonance width in magnetic field). Recent discovery of orbital Feshbach rsonance (OFR) in alkali earth $^{173}{\rm Yb}$ atoms has opened a door for a new class of resonance system \cite{Zhang,Munich,Florence}. A lot of exciting physics, for example, the massive Leggett mode, is hopefully to be observed in such systems \cite{He}. More recent studies have shown that the superfulid transition temperature of Fermi gas across such resonances behaves similar to that across a narrow resonance while the very large value of $\Delta B$ allows great controllability in many-body system \cite{Xu}. So one natural question is how the low temperature properties behave in such systems. We address this question by investigate polaronic and molecular states in this paper. Similar problems have been studied by Chen {\it et al} with a variational method \cite{Chen}. We will show that such a variational method is equivalent to the $T$-matrix approximation. Furthermore, we will include one particle-hole excitation in the molecular state and get significant correction to the polaron to molecule transition point.

The OFR consists of an orbital electronic ground state $|g\rangle$ and an excited clock state $|e\rangle$ with the nuclear spin-$\up$ or $\dn$ \cite{Zhang,Munich,Florence}. This system has an open channel $|g\dn+e\up\rangle$ and a close channel $|g\up+e\dn\rangle$ with the energy shift between this two channels $\delta$ tuned by the magnetic field. We consider here a single $|e\up\rangle$ impurity immersed in a Fermi sea $|FS_{g\dn}\rangle$ of $|g\dn\rangle$ atoms. The coupling between these two channels cause the atoms to convert between the open and close channels. Different from the magnetic Feshbach resonances, the OFR can have two kinds of particle-hole and pairing contribution in the polaronic and molecular state, which will be investigated in the following.

This paper is organized as follows. First, we look at the polaron effective mass and quasi-particle residue, the results of which show narrow resonance behaviour in the OFR. The particle-hole and pairing contribution from the open and close channels in the polaronic and molecular state are also considered. Then we investigate the polaron to molecule transition in this system. The transition point is found to move toward the BCS side when increasing the particle density. At last, we give our conclusions.

\section{Polaronic state}
\begin{figure}
  \centering
  \includegraphics[width=0.45\textwidth]{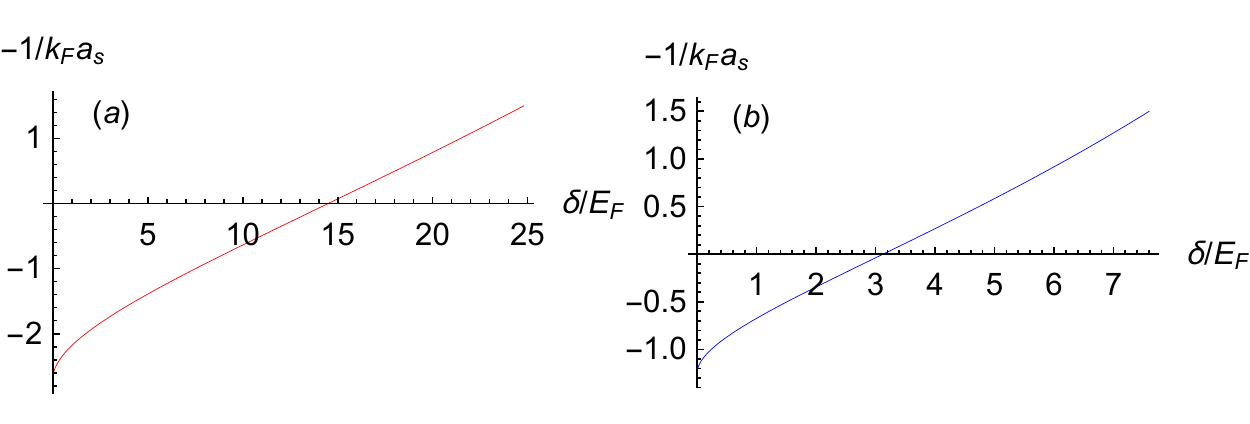}\\
  \caption{The $s$-wave scattering length $a_s$ as a function of detuning $\delta$. The red and blue lines in (a) and (b) are for single-component densities $n=k_F^3/(6\pi^2)=0.5\times10^{13}{\rm cm}^{-3}$ and $5\times10^{13}{\rm cm}^{-3}$ respectively.}
  \label{fig:fig1}
\end{figure}

We consider an alkali earth system with a Fermi sea of $|g\dn\rangle$ atoms and a single $|e\up\rangle$ impurity. The Hamiltonian can be written as
\begin{align}
H=H_0+V_0+V_1,
\end{align}
where
\begin{align}
H_0&=\sum_{\vect{k}}\varepsilon_{\vect{k}}a_{{\vect{k}},e\up}^\dagger a_{{\vect{k}},e\up}+\sum_{\vect{k}}(\varepsilon_{\vect{k}}-\mu_F)a_{{\vect{k}},g\dn}^\dagger a_{{\vect{k}},g\dn}\nonumber\\
&+\sum_{\vect{k}}\left(\varepsilon_{\vect{k}}+\delta/2\right)\left(a_{{\vect{k}},e\dn}^\dagger a_{{\vect{k}},e\dn}+a_{{\vect{k}},g\up}^\dagger a_{{\vect{k}},g\up}\right)\nonumber
\end{align}
is the non-interaction Hamiltonian for open and close channels with $\varepsilon_{\vect{k}}=\hbar^2\vect{k}^2/(2m)$ and $\mu_F=E_F$ is the chemical potential of $|g\dn\rangle$ atoms. The interaction part is
\begin{align}
V_0=&g_0\sum_{\vect{kpq}}\left(a_{{\vect{k}}/2+{\vect{q}},e\up}^\dagger a_{{\vect{k}}/2-{\vect{q}},g\dn}^\dagger a_{{\vect{k}}/2-{\vect{p}},g\dn} a_{{\vect{k}}/2+{\vect{p}},e\up}\right.\nonumber\\
&\left.+a_{{\vect{k}}/2+{\vect{q}},e\dn}^\dagger a_{{\vect{k}}/2-{\vect{q}},g\up}^\dagger a_{{\vect{k}}/2-{\vect{p}},g\up} a_{{\vect{k}}/2+{\vect{p}},e\dn}\right),\nonumber\\
V_1=&g_1\sum_{\vect{kpq}}\left(a_{{\vect{k}}/2+{\vect{q}},e\up}^\dagger a_{{\vect{k}}/2-{\vect{q}},g\dn}^\dagger a_{{\vect{k}}/2-{\vect{p}},g\up} a_{{\vect{k}}/2+{\vect{p}},e\dn}+{\rm H.c.}\right)\nonumber
\end{align}
with $g_{0}$ and $g_{1}$ gover the inter and intra coupling between open and close channels. They are connected to the spin singlet and triplet interaction strength $g_+$ and $g_-$ as $g_0=(g_++g_-)/2$ and $g_1=(g_--g_+)/2$ with the renormalization relations $1/g_\pm=m/(4\pi\hbar^2a_\pm)-\sum_{\vect{k}}{1/(2\epsilon_{\vect{k}})}$. Here $\delta$ is the energy shift between the two channels which can be tunned by the magnetic field. The effective two-body $s$-wave scattering length $a_s=[-a_0+\sqrt{m\delta/\hbar^2}(a_0^2-a_1^2)]/(a_0\sqrt{m\delta/\hbar^2}-1)$ with the scattering length of inter and intra interaction $a_0=(a_++a_-)/2$, $a_1=(a_--a_+)/2$, and the resonance position at detuning $\delta=\hbar^2/ma_0^2$ \cite{Zhang}. For the following comparison, we plot $a_s$ as a function of detuning $\delta$ for single-component particle densities $n=k_F^3/(6\pi^2)=0.5\times10^{13}{\rm cm}^{-3}$ and $5\times10^{13}{\rm cm}^{-3}$ in Fig. \ref{fig:fig1}. Hereafter we choose the experimental parameter $a_+=1900a_0$ and $a_-=200a_0$ for $^{173}{\rm Yb}$. \cite{Munich,Florence}.

The single impurity can form a polaronic state with the surrounding particles, exciting particle-hole excitations in this system. In a many-body calculation, the polaron energy corresponds to the pole of $G(\vect{p},\omega)$ at $\vect{p}=0$, i.e., $E=\Re[\Sigma(0,E)]$, where $G(\vect{p},\omega)=1/[\omega-\varepsilon_{\vect{p}}-\Sigma({\vect{p}},\omega)]$ is the single particle propagator of the impurity and $\Sigma({\vect{p}},\omega)$ is the self-energy. Basically, one can expand $\Sigma(\vect{p},\omega)$ in the number of particle-hole excitations. As noticed in \cite{polaron4}, the one particle-hole approximation in the polaronic state corresponds to the ladder self-energy
\begin{align}
\Sigma({\vect{p}},\omega)=\int \frac{d{\vect{q}}}{(2\pi)^3}\Theta({\vect{q}}-\vect{k}_F)T^{oo}({\vect{p}}+{\vect{q}},\omega+\varepsilon_{\vect{q}}),
\end{align}
where
\begin{align}
T^{oo}({\vect{q}},E+\varepsilon_{\vect{q}})\equiv\sum_{q<k_F}\frac{g_0-\Delta\chi^c}{1-g_0\chi+\Delta\chi^o\chi^c}
\label{eq:tmatrix}
\end{align}
is the scattering ladder $T$-matrix for open channel collisions defined in \cite{Xu}, with $\Delta=g_0^2-g_1^2$, $\chi=\chi^o+\chi^c$ and
\begin{align}
\chi^o&=\chi^o({\vect{q}},E+\varepsilon_{\vect{q}})\equiv\sum_{k>k_F}\frac{1}{E+\varepsilon_{\vect{q}}-\varepsilon_{{\vect{q}}-{\vect{k}}}-\varepsilon_{\vect{k}}},\\
\chi^c&=\chi^c({\vect{q}},E+\varepsilon_{\vect{q}})\equiv\sum_{\vect{p}}\frac{1}{E+\varepsilon_{\vect{q}}-\varepsilon_{{\vect{q}}-{\vect{p}}}-\varepsilon_{\vect{p}}-\delta}
\end{align}
are the particle-particle bubbles for open and close channels respectively. For convenience, we have defined $E=\mu-\mu_F$ as the relative polaron energy. So we find the polaron energy in one particle-hole excitation approximation as
\begin{align}
E=\sum_{q<k_F}T^{oo}({\vect{q}},E+\varepsilon_{\vect{q}}).
\label{eq:polaronE}
\end{align}
Here we see our $T$-matrix approximation leads to exactly the same result with the variational method carried out by Chen {\it et al } \cite{Chen}. Since we are interested in the ground state of the system, we will only focus on the polarons from the lower branch in the calculations.

Expanding the self energy $\Sigma({\vect{p}},\omega)$ to the second order in ${\vect{p}}$ around the polaron pole we get the dispersion relation for the polaron with the effective mass
\begin{align}
m^*=\frac{m}{Z}\left[1+\frac{\partial\Re[\Sigma({\vect{p}},E)]}{\partial\varepsilon_{\vect{p}}}\bigg|_{\vect{p}=0}\right]^{-1},
\end{align}
where $Z=|A|^2=[1-\partial_\omega\Sigma(0,\omega)|_{\omega=E}]^{-1}$ is the quasi-particle residue. The particle-hole contribution for open and close channel are represented respectively
\begin{align}
\sum_{q<k_F<k}|B_{\vect{qk}}|^2=-Z\sum_{q<k_F}\frac{\partial T^{oo}_{\vect{p}}}{\partial\chi^o_{\vect{p}}}\frac{\partial\chi^o_{\vect{p}}}{\partial\varepsilon_{\vect{p}}}\bigg|_{\vect{p}=0}, \\
\sum_{q<k_F,{\vect{p}}}|C_{\vect{qp}}|^2=-Z\sum_{q<k_F}\frac{\partial T^{oo}_{\vect{p}}}{\partial\chi^c_{\vect{p}}}\frac{\partial\chi^c_{\vect{p}}}{\partial\varepsilon_{\vect{p}}}\bigg|_{\vect{p}=0},
\end{align}
where $T^{oo}_{\vect{p}}=T^{oo}({\vect{p}}+{\vect{q}},E+\varepsilon_{\vect{q}})$ and $\chi^{o/c}_{\vect{p}}=\chi^{o/c}({\vect{p}}+{\vect{q}},E+\varepsilon_{\vect{q}})$.

\begin{figure}
  \centering
  \includegraphics[width=0.45\textwidth]{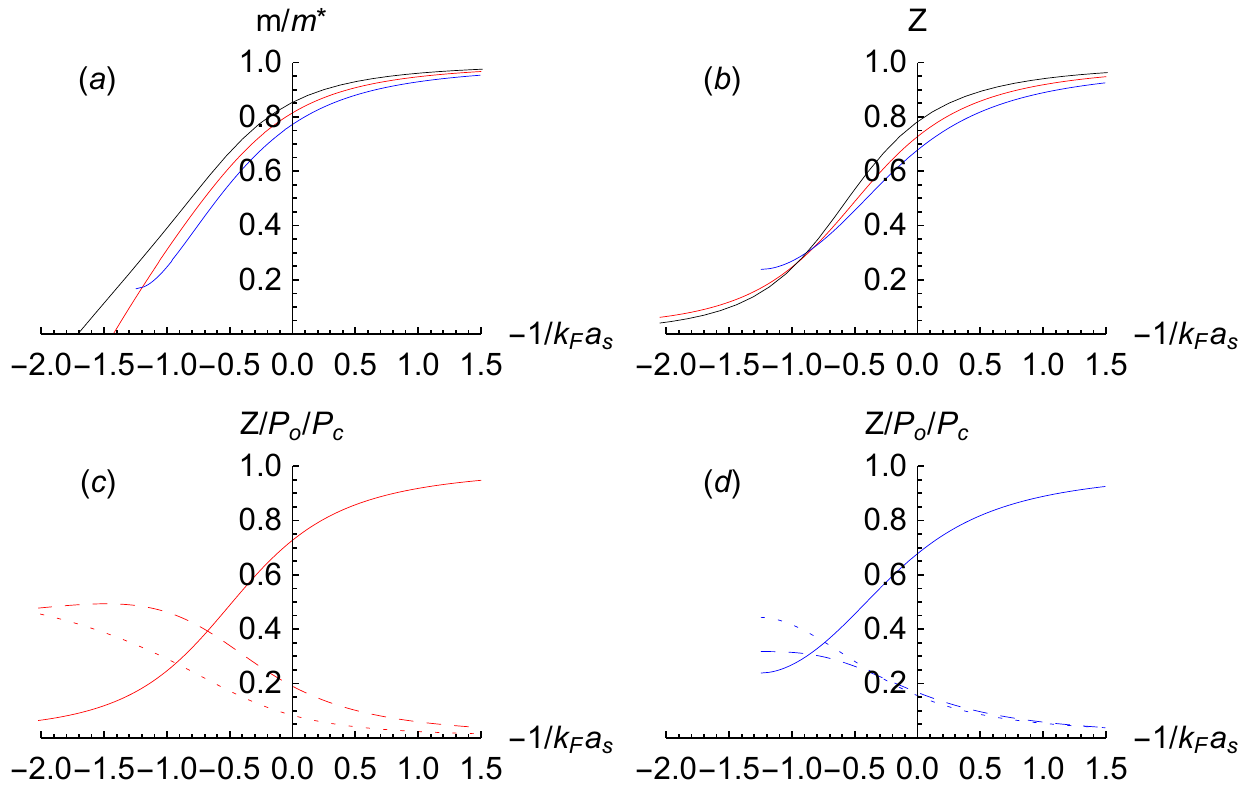}\\
  \caption{(a-b): The inverse effective mass $m/m^*$ and quasi-particle residue $Z$ as a function of interaction strength for different densities. (c-d): the quasi-particle residue $Z$ (solid lines) and open and close channel contribution $P_o=\sum_{q<k_F<k}|B_{\vect{qk}}|^2$ (dashed lines), $P_c=\sum_{q<k_F,{\vect{p}}}|C_{\vect{qp}}|^2$ (dotted lines) for different densities. The black lines are the single-channel limit. The red and blue colors are defined as in Fig. \ref{fig:fig1}.}
  \label{fig:fig2}
\end{figure}

We show in Fig. \ref{fig:fig2} the effective mass, quasi-particle residue, and particle-hole excitation contributions for different particle densities. For comparison, we also plot the zero-density single-channel limit results as black lines in Fig. \ref{fig:fig2}(a) and (b). In low-energy scattering one can expand the $s$-wave phase shift $\delta_0$ as $-1/a_s(k)\equiv k\cot\delta_0\approx-1/a_s+r_0k^2/2$, where $r_0$ is the effective range. Since $r_0$ is negative here, as $k$ increases, one needs a larger $-1/a_s$ to produce the same energy-dependent scattering length $a_s(k)$ \cite{Xu}. This will lead to the shift of inverse effective mass and quasi-particle residue toward the BCS side as one increases the particle densities, which are similar as previous studies on narrow Feshbach resonances \cite{Massignan}. In the deep BEC regime at around detuning $\delta=0$, we see a discrepancy of this shift. We attribute this to the failure of effective range expansion and a high-order correction beyond effective range may be needed in the phase shift. As illustrated in Fig. \ref{fig:fig2}(c) and (d) the quasi-particle residue dominates in the deep BCS regime. This is because the particle-hole contribution from close channel is not energetically favourable due to a large detuning at deep BCS regime, and the open channel contribution with pairing around Fermi energy has higher energy than the quasi-particle contribution. When go to the deep BEC side, as $\delta$ decreases to zero, the polaron reduces its energy by pairing at zero energy in the close channel, resulting the close channel particle-hole contribution as the dominant term.

\section{Dressed molecular state and polaron to molecule transition}
\begin{figure}
  \centering
  \includegraphics[width=0.48\textwidth]{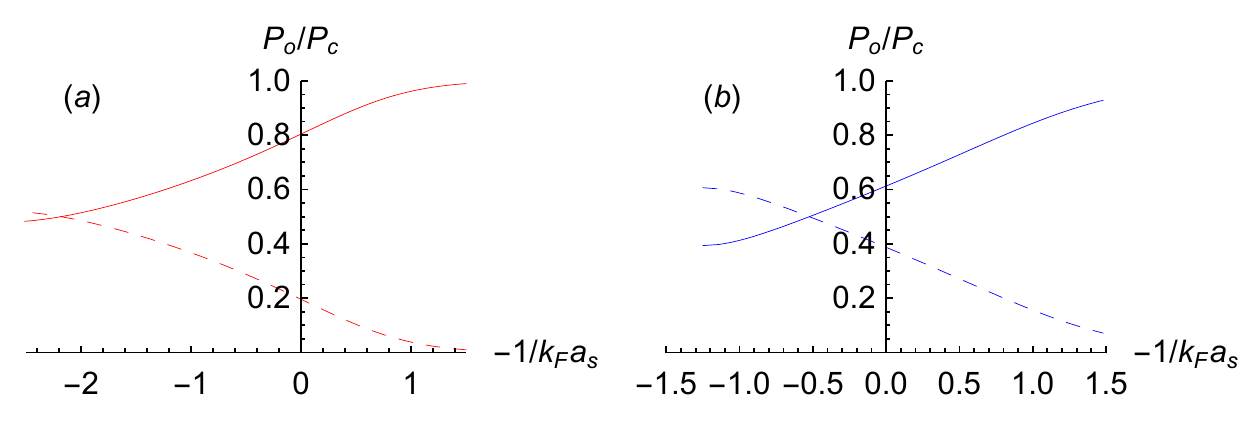}\\
  \caption{The open and close channel contribution $P_o=\sum_{k>k_F}|B_{\vect{k}}|^2$ (solid lines), $P_c=\sum_{\vect{p}}|C_{\vect{p}}|^2$ (dashed lines) in a molecular state. The red and blue colors are defined as in Fig. \ref{fig:fig1}.}
  \label{fig:fig3}
\end{figure}

The single impurity can also form a dressed molecular state by pairing in the open or close channel. Chen {\it et al} have considered such dressed molecular state without including particle-hole excitations \cite{Chen}, and they find the dimer energy corresponds to the Thouless pole of our $T$-matrix $T^{oo}$ defined in Eq. (\ref{eq:tmatrix}) for $\vect{q}=0$, i.e.,
\begin{align}
1-g_0\chi_0+\Delta\chi^o_0\chi^c_0=0,
\end{align}
where $\chi^{o/c}_0=\chi^{o/c}(0,\omega)|_{\omega=E}$, $\chi_0=\chi^o_0+\chi^c_0$, and we have defined $E=\mu-\mu_F$ as the relative dimer energy. Similarly, the pairing from open and closed channel contributions can be written as
\begin{align}
\sum_{k>k_F}|B_{\vect{k}}|^2=\frac{g_1^2\partial_\omega\chi^o_0}{g_1^2\partial_\omega\chi^o_0+(g_0-\Delta\chi^o_0)^2\partial_\omega\chi^c_0},\\
\sum_{\vect{p}}|C_{\vect{p}}|^2=\frac{(g_0-\Delta\chi^o_0)^2\partial_\omega\chi^c_0}{g_1^2\partial_\omega\chi^o_0+(g_0-\Delta\chi^o_0)^2\partial_\omega\chi^c_0}.
\end{align}
We show these two contributions for different particle densities in Fig. \ref{fig:fig3}. At the deep BCS regime where the detuning $\delta$ is large, the open channel molecule is energetically favourable and dominates. Similar like the polaron case, when go to the BEC side, the detuning $\delta$ becomes small and this energy detuning is overcome by the energy gain from pairing at Fermi sea in the open channel to zero energy pairing in the close channel. So there will be a crossover from open channel to close channel contributions. We see the cross point moves toward the BCS side for increasing particle densities in Fig. \ref{fig:fig3}.

Based on past experience in the single impurity problem \cite{Qi}, the particle-hole excitation has significant influence on the location of polaron to molecule transition point, thus to get the transition point it is necessary to include one particle-hole excitation in the molecular state. In this case, the molecular state wavefunction can be written as
\begin{align}
|\Psi_M\rangle&=\bigg(\sum_{\vect{k}}{'}B_{\vect{k}}a_{\vect{k},e\up}^\dagger a_{-\vect{k},g\dn}^\dagger+\sum_{{\vect{p}}}C_{{\vect{p}}}a_{{\vect{p}},e\dn}^\dagger a_{-{\vect{p}},g\up}^\dagger\nonumber\\
&+\sum_{\vect{kk'q}}{'}B_{\vect{k}'\vect{kq}}a_{\vect{q-k-k}',e\up}^\dagger a_{\vect{k}',g\dn}^\dagger a_{\vect{k},g\dn}^\dagger a_{\vect{q},g\dn}\nonumber\\
&+\sum_{\vect{pkq}}{'}C_{\vect{pkq}}a_{\vect{q-k-p},e\dn}^\dagger a_{\vect{p},g\up}^\dagger a_{\vect{k},g\dn}^\dagger a_{\vect{q},g\dn}\bigg)|FS_{g\dn}\rangle,
\end{align}
where the prime means the summation is restricted to $k (k')>k_F$, and $q<k_F$. The last two terms on the right include the particle-hole contributions to open and close channel molecules.
Similar like previous sections, the variation of $\langle\Psi_M|H-\mu|\Psi_M\rangle$ gives the following coupled equations
\begin{align}
E_{\vect{k}}B_{\vect{k}}=&g_0\sum_{\vect{k}}{'}B_{\vect{k}}+g_1\sum_{\vect{p}}C_{\vect{p}}-2g_0\sum_{\vect{k'q}}{'}B_{\vect{k'kq}}\nonumber\\
&-g_1\sum_{\vect{pq}}{'}C_{\vect{pkq}},\\
E_{\vect{p}}C_{\vect{p}}=&g_0\sum_{\vect{p}}C_{\vect{p}}+g_1\sum_{\vect{k}}{'}B_{\vect{k}},\\
E_{\vect{k'kq}}B_{\vect{k'kq}}=&g_0\sum_{\vect{k''}}{'}(B_{\vect{k''kq}}-B_{\vect{k''k'q}})-\frac{g_0}{2}(B_{\vect{k}}-B_{\vect{k'}})\nonumber\\
&+\frac{g_1}{2}\sum_{\vect{p}}(C_{\vect{pkq}}-C_{\vect{pk'q}}),\\
E_{\vect{pkq}}C_{\vect{pkq}}=&g_0\sum_{\vect{p'}}C_{\vect{p'kq}}-g_1B_{\vect{k}}+2g_1\sum_{\vect{k'}}{'}B_{\vect{k'kq}},
\end{align}
where $E_{\vect{k}}=E-2\ep_{\vect{k}}$, $E_{\vect{p}}=E-\delta-2\ep_{\vect{p}}$, $E_{\vect{k'kq}}=E-\ep_{\vect{q}-\vect{k}-\vect{k'}}-\ep_{\vect{k'}}-\ep_{\vect{k}}+\ep_{\vect{q}}$, and $E_{\vect{pkq}}=E-\delta-\ep_{\vect{q}-\vect{k}-\vect{p}}-\ep_{\vect{p}}-\ep_{\vect{k}}+\ep_{\vect{q}}$. Here $E=\mu-\mu_F$ is the relative dimer energy.
By defining $\eta=g_-/2(\sum_{\vect{k}}{'}B_{\vect{k}}+\sum_{\vect{p}}C_{\vect{p}})$, $\xi=g_+/2(\sum_{\vect{k}}{'}B_{\vect{k}}-\sum_{\vect{p}}C_{\vect{p}})$, $\eta_{\vect{kq}}=2g_0\sum_{\vect{k}'}{'}B_{\vect{k}'\vect{kq}}+g_1\sum_{\vect{p}}C_{\vect{pkq}}$, and $\xi_{\vect{kq}}=2g_1\sum_{\vect{k'}}{'}B_{\vect{k'kq}}+g_0\sum_{\vect{p}}C_{\vect{pkq}}$, the coupled equations can be re-written as
\begin{align}
E_{\vect{k}}B_{\vect{k}}&=\eta+\xi-\sum_{{\vect{q}}}{'}\eta_{{\vect{kq}}},\label{eq23}\\
E_{\vect{p}}C_{\vect{p}}&=\eta-\xi,\label{eq24}\\
E_{{\vect{k}}'{\vect{kq}}}B_{{\vect{k}}'{\vect{kq}}}&=\frac{1}{2}(\eta_{{\vect{kq}}}-\eta_{{\vect{k}}'{\vect{q}}})-\frac{g_0}{2}(B_{\vect{k}}-B_{{\vect{k}}'}),\label{eq25}\\
E_{{\vect{pkq}}}C_{{\vect{pkq}}}&=\xi_{{\vect{kq}}}-g_1B_{\vect{k}}.\label{eq26}
\end{align}
From Eq. (\ref{eq23}) and (\ref{eq24}) we have
\begin{align}
\eta&=-\frac{g_--g_+g_-\chi^c_0}{1-g_0\chi_0+\Delta\chi^o_0\chi^c_0}\sum_{\vect{kq}}{'}\frac{\eta_{\vect{kq}}}{2E_{\vect{k}}},\\
\xi&=-\frac{g_+-g_+g_-\chi^c_0}{1-g_0\chi_0+\Delta\chi^o_0\chi^c_0}\sum_{\vect{kq}}{'}\frac{\eta_{\vect{kq}}}{2E_{\vect{k}}},
\end{align}
where $\chi^o_0=\chi^o(0,E)=\sum_{\vect{k}}{'}1/E_{\vect{k}}$, $\chi^c_0=\chi^c(0,E)=\sum_{\vect{p}}1/E_{\vect{p}}$, $\chi_0=\chi^o_0+\chi^c_0$, and $\Delta=g_0^2-g_1^2$. Plug $\eta$ and $\xi$ into Eq. (\ref{eq23}) we have
\begin{align}
B_{\vect{k}}=-\sum_{\vect{k'q'}}{'}\frac{\eta_{\vect{k'q'}}}{\gamma E_{\vect{k}}E_{\vect{k'}}}-\sum_{\vect{q'}}{'}\frac{\eta_{\vect{kq'}}}{E_{\vect{k}}},
\end{align}
where $\gamma=1/T^{oo}(0,E)=1/a-\chi^o_0$ with $a=(g_0-\Delta\chi^c_0)/(1-g_0\chi^c_0)$.
From Eq. (\ref{eq25}) and (\ref{eq26}) we have
\begin{align}
\eta_{\vect{kq}}&=\frac{1}{1-g_0\chi_{\vect{kq}}^o}\left[g_1\chi_{\vect{kq}}^c\xi_{\vect{kq}}-(g_0^2\chi_{\vect{kq}}^o+g_1^2\chi_{\vect{kq}}^c)B_{\vect{k}}\right.\nonumber\\
&\left.-g_0\sum_{\vect{k'}}{'}\frac{\eta_{\vect{k'q}}}{E_{\vect{k'kq}}}+g_0^2\sum_{\vect{k'}}{'}\frac{B_{\vect{k'}}}{E_{\vect{k'kq}}}\right],\label{eq31}\\
\xi_{\vect{kq}}&=\frac{1}{1-g_0\chi_{\vect{kq}}^c}\left[g_1\chi_{\vect{kq}}^o\eta_{\vect{kq}}-g_0g_1\chi_{\vect{kq}}B_{\vect{k}}\right.\nonumber\\
&\left.-g_1\sum_{\vect{k'}}{'}\frac{\eta_{\vect{k'q}}}{E_{\vect{k'kq}}}+g_0g_1\sum_{\vect{k'}}{'}\frac{B_{\vect{k'}}}{E_{\vect{k'kq}}}\right],\label{eq32}
\end{align}
where $\chi_{{\vect{kq}}}^o=\chi^o({\vect{q}}-\vect{k},E+\epsilon_{\vect{q}}-\epsilon_{\vect{k}})=\sum_{{\vect{k}}'}{'}1/E_{{\vect{k}}'{\vect{kq}}}$, $\chi_{{\vect{kq}}}^c=\chi^c({\vect{q}}-\vect{k},E+\epsilon_{\vect{q}}-\epsilon_{\vect{k}})=\sum_{\vect{p}}1/E_{{\vect{pkq}}}$, and $\chi_{{\vect{kq}}}=\chi_{{\vect{kq}}}^o+\chi_{{\vect{kq}}}^c$.
Plug $\xi_{\vect{kq}}$ in Eq. (\ref{eq32}) into Eq. (\ref{eq31}) we have
\begin{align}
&(1-g_0\chi_{\vect{kq}}^o)\eta_{\vect{kq}}\nonumber\\
=&\frac{g_1^2\chi_{\vect{kq}}^o\chi_{\vect{kq}}^c}{1-g_0\chi_{\vect{kq}}^c}\eta_{\vect{kq}}-\left[\frac{g_0-\Delta\chi_{\vect{kq}}^c}{1-g_0\chi_{\vect{kq}}^c}(1+g_0\chi_{\vect{kq}}^o)-g_0\right]B_{\vect{k}}\nonumber\\
&-\frac{g_0-\Delta\chi_{\vect{kq}}^c}{1-g_0\chi_{\vect{kq}}^c}\sum_{\vect{k'}}{'}\frac{\eta_{\vect{k'q}}}{E_{\vect{k'kq}}}+\frac{g_0(g_0-\Delta\chi_{\vect{kq}}^c)}{1-g_0\chi_{\vect{kq}}^c}\sum_{\vect{k'}}{'}\frac{B_{\vect{k'}}}{E_{\vect{k'kq}}}\nonumber.
\end{align}

So by plunging $B_{\vect{k}}$ into above equation and after some simplification we get the final integral equation
\begin{align}
\gamma_E\eta_{\vect{kq}}=\sum_{\vect{k'q'}}{'}\frac{\eta_{\vect{k'q'}}}{\gamma E_{\vect{k'}}E_{\vect{k}}}+\sum_{\vect{q'}}{'}\frac{\eta_{\vect{kq'}}}{E_{\vect{k}}}-\sum_{\vect{k'}}{'}\frac{\eta_{\vect{k'q}}}{E_{\vect{k'kq}}},\label{eq:integral}
\end{align}
where $\gamma_E=1/T^{oo}({\vect{q}}-\vect{k},E+\epsilon_{\vect{q}}-\epsilon_{\vect{k}})=1/a_E-\chi_{{\vect{kq}}}^o$, with $a_E=(g_0-\Delta\chi_{{\vect{kq}}}^c)/(1-g_0\chi_{{\vect{kq}}}^c)$. We see the scattering length corresponds to $a_s=a(E\to0)=a_E({\vect{k}}\to0,{\vect{q}}\to0,E\to0)$.

\begin{figure}[h]
  \centering
  \includegraphics[width=0.45\textwidth]{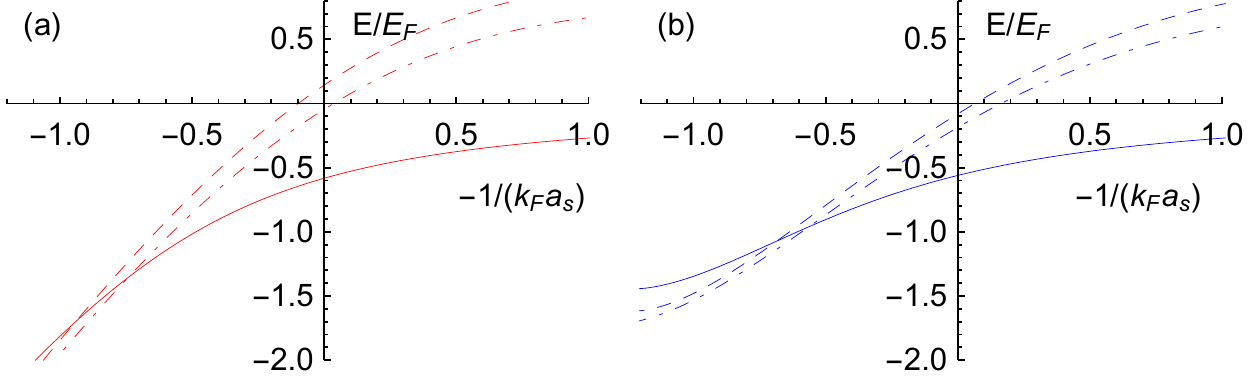}\\
  \caption{The polaron and molecule energy as a function of interaction strength for different densities. The solid lines are polaron energies including one particle-hole excitation, while the dashed/dash-dotted lines are molecule energies without/with one particle-hole excitation. Here the red and blue colors are defined as in Fig. \ref{fig:fig1}.}
  \label{fig:fig4}
\end{figure}

\begin{figure}[h]
  \centering
  \includegraphics[width=0.48\textwidth]{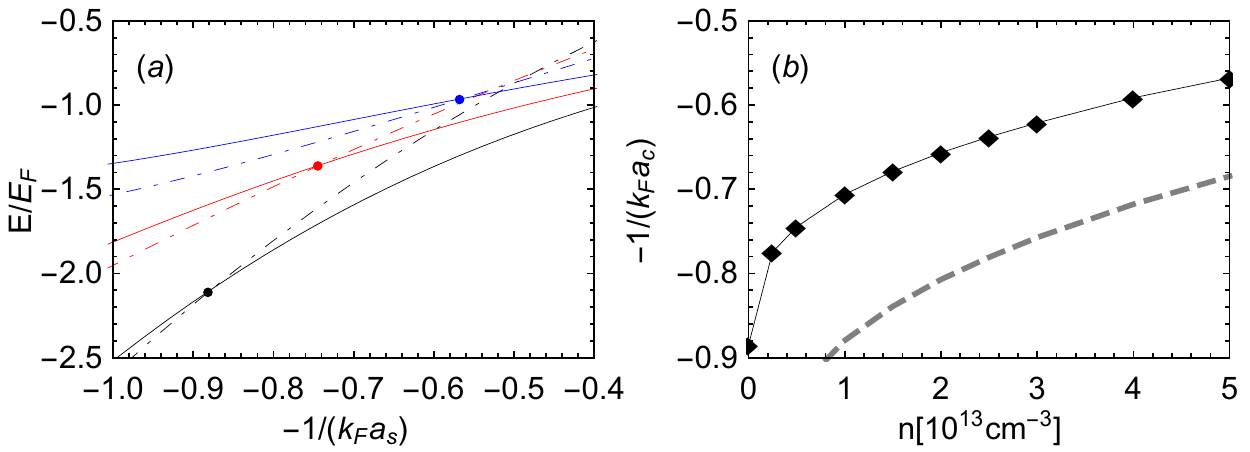}\\
  \caption{The polaron to molecule transition for different densities. (a) The solid/dash-dotted lines are polaron/molecule energies including one particle-hole excitation, with the red and blue colors are defined as in Fig. \ref{fig:fig1} and the black lines are results of single-channel limits. (b) The critical interaction strength as a function of particle density. For comparing, we also plot the results without particle-hole excitation in the molecular state as the dashed line in (b).}
  \label{fig:fig5}
\end{figure}

This integral equation (\ref{eq:integral}) can be solved numerically and the molecular state energy is got when the Fredholm determinant of the kernel vanishes \cite{1PH1,1PH2,1PH3}. We solve Eq. (\ref{eq:integral}) with $q=0$ as an approximation. We show the polaron and molecule energy for two different densities in Fig. \ref{fig:fig4}. For comparison, we also show the molecule energy without particle-hole excitations as dashed lines. The polaron to molecule transition is shown in Fig. \ref{fig:fig5} for increasing particle densities. The inclusion of one particle-hole excitation in the molecular state makes significant correction to the transition point, as shown in Fig. \ref{fig:fig5}(b), where the dashed line indicates the result without particle-hole excitation in the molecular state. In the zero-density limit, the energy dependence in scattering amplitude can be neglected and our results recovers that of a single-channel mode as it should be. As the density increases, the polaron to molecule transition point goes toward the BCS side for increasing densities, which is similar to previous studies in narrow Feshbach resonances \cite{Qi,Castin,Massignan}.

\section{Conclusions}
To conclude, we have studied the polaronic and dressed molecular states in a OFR,  considering a single $|e\up\rangle$ atom immersed in a $|g\dn\rangle$ Fermi sea. Our $T$-matrix approximation is found to be equivalent to the variational wave function method. We find significant correction to the polaron to molecule transition point by including one particle-hole excitation in the molecular state. The transition point moves toward the BCS side for increasing particle densities. The effective mass and quasi-particle residue is also consistent with the finite effective range behaviour. Different from single-channel Feshbach resonances, in the OFR we have two pairing contributions, corresponding to the open and close channels. The crossover from open to close channel pairing moves toward the BCS side for increasing particle densities at both polaronic and molecular states.

Note that our results only include one particle-hole excitation. Studies in narrow Feshbach resonances have shown that the inclusion of a second particle-hole excitation only makes slight changes to the system \cite{Qi}.

\section*{Acknowledgments}
We are grateful to Matthias Punk for his help on the numerical calculation. JX is supported by NSFC (No. 11504021), and FRFCU (No. FRF-TP-17-023A2). RQ is supported by NSFC (No. 11774426), the Fundamental Research Funds for the Central Universities, and the Research Funds of Renmin University of China under Grants No. 15XNLF18 and No. 16XNLQ03.

\section*{Author contribution}
JX performed the numerical calculation and RQ provided the method. Both authors were involved in preparing the manuscript.

\end{document}